# Evidence for Two-dimensional Weyl Fermions in Air-Stable Monolayer PtTe$_{1.75}$

Zhihao Cai,[1,2,#] Haijun Cao,[1,2,#] Haohao Sheng[1,2,], Xuegao Hu,[1,2] Zhenyu Sun,[1,2] Qiaoxiao Zhao,[1,2] Jisong Gao,[1,2] Shin-ichiro Ideta,[3] Kenya Shimada,[3] Jiawei Huang,[4] Peng Cheng,[1,2] Lan Chen,[1,2,4*] Yugui Yao,[5] Sheng Meng,[1,2,4,6*] Kehui Wu,[1,2,4,6*] Zhijun Wang,[1,2] Baojie Feng[1,2,4,6*]

[1]*Institute of Physics, Chinese Academy of Sciences, Beijing, 100190, China*

[2]*School of Physical Sciences, University of Chinese Academy of Sciences, Beijing, 100049, China*

[3]*Hiroshima Synchrotron Radiation Center, Hiroshima University, Higashi-Hiroshima 739-0046, Japan*

[4]*Songshan Lake Materials Laboratory, Dongguan, Guangdong, 523808, China*

[5]*Centre for Quantum Physics, Key Laboratory of Advanced Optoelectronic Quantum Architecture and Measurement (MOE), School of Physics and Beijing Key Lab of Nanophotonics Ultrafine Optoelectronic Systems, School of Physics, Beijing Institute of Technology, Beijing 100081, China*

[6]*Interdisciplinary Institute of Light-Element Quantum Materials and Research Center for Light-Element Advanced Materials, Peking University, Beijing, 100871, China*

[#]These authors contributed equally to this work.

*Corresponding author. E-mail: lchen@iphy.ac.cn; smeng@iphy.ac.cn; khwu@iphy.ac.cn; bjfeng@iphy.ac.cn.

**Key Words: 2D Weyl fermions, PtTe$_{1.75}$, ARPES, first-principles calculations**

## Abstract

The Weyl semimetals represent a distinct category of topological materials wherein the low-energy excitations appear as the long-sought Weyl fermions. Exotic transport and optical properties are expected because of the chiral anomaly and linear energy-momentum dispersion. While three-dimensional Weyl semimetals have been successfully realized, the quest for their two-dimensional (2D) counterparts is ongoing. Here, we report the realization of 2D Weyl fermions

**in monolayer PtTe$_{1.75}$, which has strong spin-orbit coupling and lacks inversion symmetry, by combined angle-resolved photoemission spectroscopy, scanning tunneling microscopy, second harmonic generation, X-ray photoelectron spectroscopy measurements, and first-principles calculations. The giant Rashba splitting and band inversion lead to the emergence of three pairs of critical Weyl cones. Moreover, monolayer PtTe$_{1.75}$ exhibits excellent chemical stability in ambient conditions, which is critical for future device applications. The discovery of 2D Weyl fermions in monolayer PtTe$_{1.75}$ opens up new possibilities for designing and fabricating novel spintronic devices.**

Topological semimetals, including Dirac, Weyl, and nodal line semimetals, are characterized by linearly dispersing bands that intersect at discrete points or extended lines [1-5]. The distinct topological electronic structures of topological semimetals give rise to a plethora of exotic properties, including high carrier mobility, Klein paradox, chiral anomaly, and extremely high magnetoresistance. These properties offer extensive potential in diverse fields, from electronics and spintronics to optoelectronic applications. Among topological semimetals, Weyl semimetals are particularly captivating [6-16]. The low-energy excitations of Weyl semimetals behave as Weyl fermions, which have been sought for nearly a century but have not been observed as fundamental particles in nature. Weyl semimetals have pairs of Weyl nodes with opposite chirality, which are like sinks and sources of Berry curvature in the momentum space. Unlike Dirac and nodal line semimetals, Weyl semimetals only need translational symmetry of the crystal, which makes them more resistant to external perturbations. To date, a variety of three-dimensional crystals have been found to host Weyl fermions, exemplified by TaAs [6-8], WTe$_2$ [9,13,14], and Co$_3$Sn$_2$S$_2$ [10-12], and they exhibit a range of exotic transport and optical properties [15-19].

Meanwhile, research on two-dimensional (2D) materials has boomed due to the ongoing trend of device miniaturization. Over recent decades, various 2D materials have been realized, exemplified by graphene, transition metal dichalcogenides, black phosphorus, and borophene, which offer a fertile ground for designing and fabricating quantum devices on the atomic scale [20-25] In addition, the properties of 2D materials are more convenient to manipulate by external strain or interfacial effects. Therefore, it is highly desirable to realize 2D Weyl semimetals for device applications. Unlike their 3D counterparts, 2D Weyl semimetals require additional symmetries to stabilize the Weyl points [26,27]. With specific symmetries, 2D Weyl semimetals can host pairs of 2D gapless Weyl cones with topological edge states that connect the projections of Weyl points, analogous to the 3D Weyl cones and Fermi arc surface states in 3D Weyl semimetals (see Fig. 1a). Because of the additional symmetry requirements, 2D Weyl semimetals can be driven into different topological phases by breaking specific symmetries, which is difficult to realize in 3D Weyl semimetal.

Recently, several candidate 2D Weyl semimetals have been theoretically predicted, including $PtCl_3$, $Cr_2C$, MFeSe (M=Tl, In, Ga), and Si/Bi heterostructure [26-30], but synthesis of these materials is challenging, which hinders their experimental investigation. Although extrinsic 2D Weyl fermion states can be induced by external stimuli, including strong magnetic field and substrate interaction [31,32], 2D materials that host intrinsic Weyl fermion states, which are promising in future spintronic devices, are still lacking.

Here, we present experimental evidence for 2D Weyl fermion states in a non-centrosymmetric 2D material, the monolayer $PtTe_{1.75}$, by combined angle-resolved photoemission spectroscopy (ARPES), scanning tunneling microscopy (STM), second harmonic generation (SHG), X-ray photoelectron spectroscopy (XPS) measurements, and first-principles calculations. The strong spin-orbit coupling (SOC) and lack of inversion symmetry give rise to three pairs of Weyl nodes in the first BZ. The linearly dispersing Weyl cones have been directly observed by systematic ARPES measurements, substantiated by first-principles calculations. In addition, monolayer $PtTe_{1.75}$ is extremely stable in various environments, including air, alcohol, and acetone, as sharp low-energy electron diffraction (LEED) patterns survive without degassing in vacuum. Our results establish monolayer $PtTe_{1.75}$ as an ideal platform to investigate the intriguing properties of Weyl fermions in the 2D limit.

Monolayer PtTe$_{1.75}$ has a non-centrosymmetric structure with a space group of P3m1, as depicted schematically in Fig. 1b. The structure of monolayer PtTe$_{1.75}$ can be understood based on the Dirac semimetal 1T-PtTe$_2$ [33] composed of one Pt layer sandwiched between two Te layers. Monolayer PtTe$_{1.75}$ can be obtained by removing one Te atom in the topmost Te layer in the 2×2 supercell of monolayer PtTe$_2$. The phonon spectrum of freestanding PtTe$_{1.75}$, shown in Fig. S1, has no imaginary frequencies, indicating the dynamic stability of monolayer PtTe$_{1.75}$.

The band structure of monolayer PtTe$_{1.75}$ is shown in Fig. 1d. Because of the strong SOC and absence of inversion symmetry, a giant Rashba splitting occurs at the Γ point, as indicated by the red circle in Fig. 1d. The giant Rashba splitting might contribute to a large spin Hall conductivity [34]. The spin-split bands are inverted, forming a twofold degenerate critical Weyl point in the M$_{100}$∗T invariant line Γ-K, as indicated by the red arrow in Fig. 1d. The symmetry protection by $[TM_{100}]^2 = 1$ [34] is different from the conventional 2D Weyl point protected by PT in spinless systems [35,36] and TC$_{2z}$ system in spinful systems [37,38]. The critical Weyl point emerges as the SOC strength approaches a critical value.

Because of the coexistence of C$_{3z}$ and M$_{100}$ symmetries, the Weyl point appear in sextuplet in the first BZ, as shown in the inset of Fig. 1d. The validity of Weyl points has also been confirmed by our Berry curvature calculations, as shown in Fig. 1e. To reveal the spin polarization of the Weyl cone, we calculated the spin texture of PtTe$_{1.75}$ with SOC, as shown in Fig. 1(f-h) and Fig. S2. Due to the coexistence of the time reversal and mirror symmetries, the S$_x$ component of the Weyl cone is required to be zero along ΓK direction and has opposite signs along the perpendicular direction. The two linearly dispersing bands of the Weyl cone have the same sign in the S$_y$ and S$_z$ components along both ΓK and the perpendicular direction passing through the Weyl point. This exotic spin texture is distinct from the helical spin polarization of Rashba splitting and Dirac cone surface states. Instead, the two Weyl cones have opposite spin polarization.

Monolayer PtTe$_{1.75}$ can be synthesized by growing Te on the Pt(111) substrate [39]. Alternatively, PtTe$_{1.75}$ quantum dots can be obtained by annealing PtTe$_2$ crystals in an ultrahigh vacuum [40]. However, previous studies were primarily focused on structural analysis, while experimental research on the electronic structure and topological properties is still lacking. To elucidate the intriguing properties of monolayer PtTe$_{1.75}$, Te atoms were deposited on Pt(111) at room temperature. At low annealing temperatures, monolayer to multilayer PtTe$_2$ will form, depending on the coverage of Te. Elevated annealing temperature will cause the desorption of excess Te atoms and the formation of periodic Te vacancies on the topmost Te layer, forming the desired monolayer PtTe$_{1.75}$ on Pt(111). Upon reaching an annealing temperature of approximately 600 K, the surface is dominated by monolayer PtTe$_{1.75}$.

In the PtTe$_{1.75}$/Pt(111) system, the Te vacancies sit at the 3×3 sites of the topmost Pt layer, forming a commensurate 3×3 superstructure with respect to Pt(111) or a 2×2 superstructure with respect to PtTe$_2$ [39]. To ascertain the vacancy sites, we calculated the adsorption energies for 8 possible configurations of the PtTe$_{1.75}$/Pt(111) system, as depicted in Fig. S3. Our calculations demonstrate that the Te vacancies are located on the top surface of PtTe$_2$ and at the bridge sites of Pt(111), as illustrated in Fig. 2a. Figure 2b shows the Brillouin Zones (BZs) of PtTe$_{1.75}$ and the Pt(111) substrate. The 2×2 vacancy superstructure with respect to PtTe$_2$ results in a kagome-like lattice for the topmost Te atoms, which is directly observed in the STM image shown in Fig. 2c. The 3×3 superstructure with respect to Pt(111) has been directly observed by low-energy electron diffraction (LEED) measurements, as shown in Fig. 2d. The sharp LEED patterns indicate the high quality and uniformity of the sample, which is suitable for ARPES measurements.

We investigated the chemical stability of monolayer PtTe$_{1.75}$ by exposing it to air and alcohol acetone mixture after growth in ultrahigh vacuum. We then transferred the sample back to ultrahigh vacuum and performed LEED and X-ray photoelectron spectroscopy (XPS) measurements. Remarkably, the LEED patterns of PtTe$_{1.75}$/Pt(111) remained intact even without degassing, as shown in Fig. S4. XPS spectra show that the Te 3d and Pt 4f peaks have negligible change in the line shape (Fig. S5). These results demonstrate the exceptional chemical stability of monolayer PtTe$_{1.75}$, comparable to graphene and MoS$_2$, which makes it possible to perform *ex-situ* optical measurements. SHG is a second-order nonlinear optical process that occurs in non-centrosymmetric crystals. As shown in Fig. 2e, an obvious spectral peak at 610 nm is observed when a femtosecond-pulsed laser at 1210 nm is incident on monolayer PtTe$_{1.75}$ (see Methods and Fig. S6a,b in the Supporting Information for more experimental details). The intensity of this spectral peak exhibits a power dependence with a slope of approximately 1.88 in the double logarithmic plot (Fig. S6c), indicating the second-order nonlinear optical nature of the process. Figure 2f shows the SHG signal with normalized intensity excited from 1225 to 1390 nm. The wavelength-dependent SHG of the PtTe$_{1.75}$ monolayer can be quantitatively analyzed through its linear optical properties and second-order nonlinear optical susceptibility (see Fig. S7 for details). These SHG results provide important evidence of inversion symmetry breaking in the PtTe$_{1.75}$ monolayer, which is crucial for the existence of Weyl fermions.

After preparing high-quality PtTe$_{1.75}$, we conducted *in-situ* ARPES measurements to study its electronic structure. The constant energy contours (CECs) of PtTe$_{1.75}$/Pt(111) and pristine Pt(111) are displayed in Fig. 3a-3d and 3e-3h, respectively. On the Fermi surface, most bands are derived from the Pt(111) substrate, except for a pocket-like feature emerging at each M point of PtTe$_{1.75}$. The spectral weight of the pocket is stronger at higher-order BZs because of the photoemission matrix element effect.

Near the Γ point of the first BZ, a faint spectral weight emerges after the growth of PtTe$_{1.75}$, which evolves into a circle-like feature at a binding energy of 0.15 eV (see Fig. 3b). At higher binding energies, the circle-like feature at the Γ point slightly shrinks, accompanied by the emergence of six nodes at approximately the halfway of Γ-K, as shown in Fig. 3c. Each node corresponds to a Weyl point predicted in freestanding PtTe$_{1.75}$. The circle shrinks to a dot-like feature at 0.9 eV and reappear at higher binding energies, as shown in Fig. 3d and 3e. The circle-like feature agrees with the inner branch of the Rashba-split bands of PtTe$_{1.75}$. It should be noted that pristine Pt(111) does not have similar bands given any energy or momentum shifts, which rules out a rigid shift of the Fermi level or the folding effects of bulk bands.

Notably, the bulk bands of Pt(111) have stronger spectral weights near the BZ boundary of Pt(111), where they overlap with the bands of PtTe$_{1.75}$. To clearly resolve the band structures of the PtTe$_{1.75}$, we focus on the band structure in the first BZ, where the bulk bands of Pt(111) have less spectral weight. Figure 4a illustrates the distribution of the six Weyl nodes and the momentum cuts along which ARPES spectra were acquired. Along the Γ-K direction, linearly dispersing bands with a cone-like shape were observed, as indicated by the black arrows in Fig. 4b. The band degenerate point is located at approximately 0.4 eV below the Fermi level, which coincides with the appearance of six nodes on the CEC at $E_B$=0.4 eV (see Fig. 3c). Along the perpendicular direction, *i.e.*, Cut AB, linearly dispersing bands with a crossing point at 0.4 eV were also observed, as shown in Fig. 4e. These results fully agree with the existence of six Weyl cones in the first BZ of PtTe$_{1.75}$.

To further confirm the 2D character of the Weyl cone in PtTe$_{1.75}$, we present ARPES spectra along the Γ-K direction measured with different photon energies, as shown in Fig. 4f-4g. The Weyl cones have negligible dispersion with photon energies, which agrees with their 2D character. In contrast, the nearby bulk bands of Pt(111) show apparent dispersion with photon energies. These results provide further evidence for the existence of 2D Weyl cones in monolayer PtTe$_{1.75}$.

As monolayer PtTe$_{1.75}$ is grown on Pt(111), the substrate-overlayer interaction is non-negligible. To clarify the influence of the substrate, we carried out first-principles calculations including the Pt(111) substrate. The structural optimization shows that monolayer PtTe$_{1.75}$ remains intact on Pt(111) with negligible buckling, as shown in Fig. 2a. Figure 4d displays the calculated band structure along the Γ-K direction. The Weyl points moved to below the Fermi level because of the electron doping from the substrate. In general, the calculated band structures agree well with our experimental results, as indicated by the black dashed lines in Fig. 4d. The survival of Weyl cones in the

PtTe$_{1.75}$/Pt(111) system suggests that the band hybridization is not strong enough to destroy the Weyl fermions, despite the significant charge transfer at the interface. The charge transfer has also been confirmed by XPS measurements [39].

Finally, we discuss the potential of isolating freestanding PtTe$_{1.75}$ for device applications. Fang et al. recently showed that annealing PtTe$_2$ crystals in ultrahigh vacuum can produce small pieces of PtTe$_{1.75}$ on the top layer [40]. Since PtTe$_2$ is a layered transition metal dichalcogenide, the top layer could be isolated by mechanical exfoliation. In addition, a similar compound, PtSe$_2$, have been successfully peeled off from the Pt(111) substrate [41]. Therefore, monolayer PtTe$_{1.75}$ might be detachable from the Pt(111) substrate using comparable methods. Our study will not only stimulate further theoretical and experimental investigations of the exotic properties of the 2D Weyl fermions but also inform ongoing research efforts in designing and fabricating novel quantum devices.

In conclusion, our combined experimental and theoretical results provide evidence for the existence of 2D Weyl fermions in monolayer PtTe$_{1.75}$. Because of the strong spin-orbit coupling and lack of inversion symmetry, monolayer PtTe$_{1.75}$ hosts a giant Rashba splitting and band inversion, giving rise to three pairs of Weyl cones. Combined ARPES measurements and first-principles calculations confirmed the presence of 2D Weyl cones. In addition, our LEED, XPS, and ARPES measurements show that monolayer PtTe$_{1.75}$ has excellent stability in various chemical environments, which is superior to most surface structures and 2D materials.

**Methods**

**Sample preparation.** Single-crystal Pt(111) was cleaned by repeated Ar ion sputtering and annealing cycles. Monolayer PtTe$_{1.75}$ was prepared by evaporating Te from a Knudsen cell onto the Pt(111) surfaces followed by annealing at the temperature of 600 K.

**ARPES experiments.** Synchrotron-based ARPES measurements were carried out at Beamline BL-1 of the Hiroshima synchrotron radiation center. During Synchrotron-based ARPES measurements, the pressure was ~1.3×10$^{-9}$ Pa, and the temperature of the samples was kept at ~15 K. ARPES experiments were also performed at our lab-based ARPES system with a SPECS PHOIBUS 150 electron energy analyzer and a helium discharge lamp (He Iα light). The base pressure during ARPES measurements was better than 2×10$^{-8}$ Pa, and the temperature of the samples was kept at ~40 K during measurements.

**STM Measurements.** STM experiments were carried out in a home-built STM system at 77 K with a base pressure better than 2×10$^{-8}$ Pa. The STM was equipped with an electrochemically etched tungsten tip and bias voltages were applied to the sample.

**SHG measurements.** Reflective optical SHG measurements were obtained using a home-built SHG system at room temperature. A tunable laser ranging from 1200 to 1400 nm was generated from an optical parametric amplifier (Orpheus, Light Conversion) pumped by a 1030-nm Yb: kGW amplified laser source (Pharos, Light Conversion), with a pulse width of approximately 200 fs and a repetition rate of 750 kHz. The excitation laser was focused using a 40x (NA=0.6) achromatic objective lens. The reflective fundamental laser and SHG signal were separated using a dichroic mirror. The SHG signal was collected and detected by spectrometers (AvaSpec-ULS) from Avantes Inc.

**Calculations.** We performed the first-principles calculations based on density functional theory (DFT) with the projector augmented wave method by using the Vienna ab initio simulation package (VASP) [42,43]. The generalized gradient approximation (GGA) with Perdew-Burke-Ernzerhof (PBE) exchange-correlation functional was adopted [44,45]. The kinetic energy cutoff was set as 400 eV for the plane-wave basis and a Γ-centered 6×6×1 $k$-point mesh in the first BZ was used. The slab included four layers of Pt substrate, monolayer PtTe$_{1.75}$ and a vacuum layer that was larger than 15 Å. The convergence conditions for the ionic relaxation loop and electronic self-consistent loop were 0.01 eV/Å and 10$^{-6}$ eV, respectively. The phonon spectra were calculated with the finite-difference method using a 3×3×1 supercell, as implemented in the Phonopy package [46].

**Associated Contents**

**Supporting Information**

The distribution of Berry curvature of freestanding monolayer $PtTe_{1.75}$; Calculated spin texture in freestanding monolayer $PtTe_{1.75}$; Total energy calculation results for the 8 possible $PtTe_{1.75}$ configurations on Pt(111); Chemical stability of $PtTe_{1.75}$/Pt(111); XPS measurements of monolayer $PtTe_{1.75}$.

**Acknowledgments**

We thank Yogendra Kumar for the help in the synchrotron ARPES experiments. This work was supported by the Beijing Natural Science Foundation (Grant No. JQ23001), the Ministry of Science and Technology of China (Grants No. 2018YFE0202700), the National Natural Science Foundation of China (Grant No. 12374197), the Strategic Priority Research Program of Chinese Academy of Sciences (Grants No. XDB33030100), and the CAS Project for Young Scientists in Basic Research (Grant No. YSBR-047). The synchrotron ARPES experiments were performed with the approval of the Proposal Assessing Committee of the Hiroshima Synchrotron Radiation Center (Proposals No. 23AG024 and No. 23AG025).

**Competing Interests**

The authors declare no competing interests.

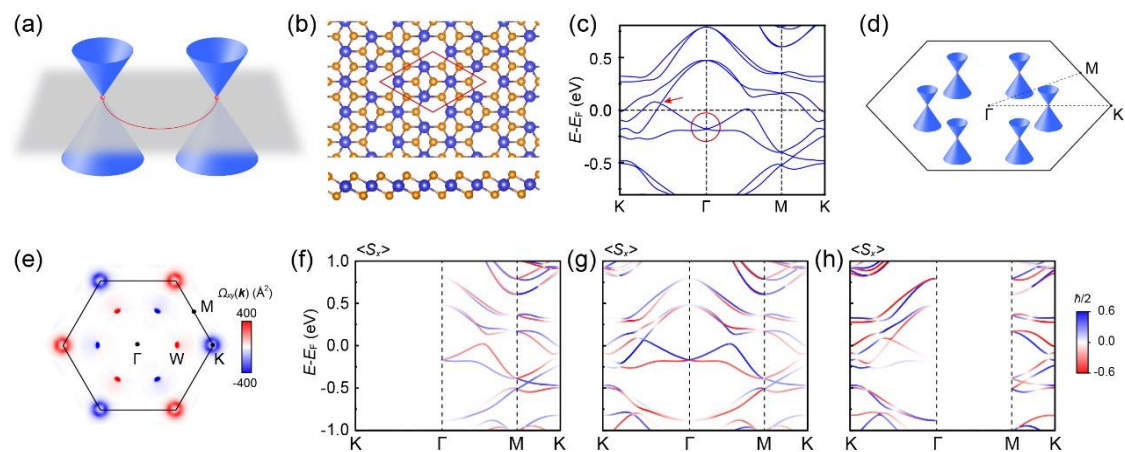

**FIG. 1: Overview of 2D Weyl fermions and monolayer PtTe$_{1.75}$.** (a) Schematic drawing of a pair 2D Weyl cones. There are topological edge states connecting the Weyl nodes, akin to the topological Fermi arcs in 3D Weyl semimetals. (b) Top view and side view of the structure model of monolayer PtTe$_{1.75}$. Orange and blue balls represent the Te and Pt atoms, respectively. The red rhombus indicates a unit cell. (c) Calculated band structure of monolayer PtTe$_{1.75}$ with SOC. The red circle indicates the Rashba splitting at the Γ point. The red arrow indicates Weyl cones between the Γ and K point. (d) The distribution of the Weyl cones in the first BZ. (e) The distribution of Berry curvature of freestanding monolayer PtTe$_{1.75}$. (f-h) Calculated spin texture in freestanding monolayer PtTe$_{1.75}$.

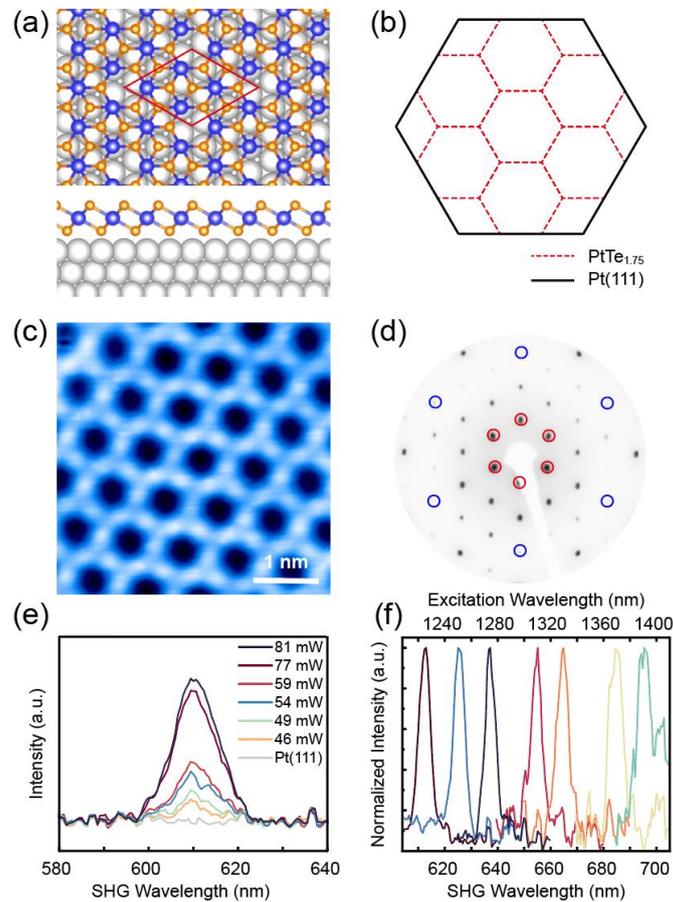

**FIG. 2: Structure and growth of monolayer PtTe$_{1.75}$ on Pt(111).** (a) Top view and side view of the structure model of monolayer PtTe$_{1.75}$ on Pt(111). Orange and blue balls represent the Te and Pt atoms in PtTe$_{1.75}$, respectively. Gray balls represent the Pt atoms in the Pt(111) substrate. The red rhombus indicates a unit cell of PtTe$_{1.75}$, which is a commensurate 3×3 superstructure with respect to Pt(111). (b) Schematic drawing of the BZs of the PtTe$_{1.75}$ (red) and Pt(111) (black). (c,d) STM topographic image and LEED patterns of PtTe$_{1.75}$/Pt(111), respectively. Blue and red circles indicate

the spots of Pt(111) and PtTe$_{1.75}$, respectively. The incident electron energy is 60 eV. (e) SHG spectra of monolayer PtTe$_{1.75}$ under excitation of 1210 nm with various powers; (f) Normalized SHG intensity with excitation laser wavelength from 1225 to 1390 nm.

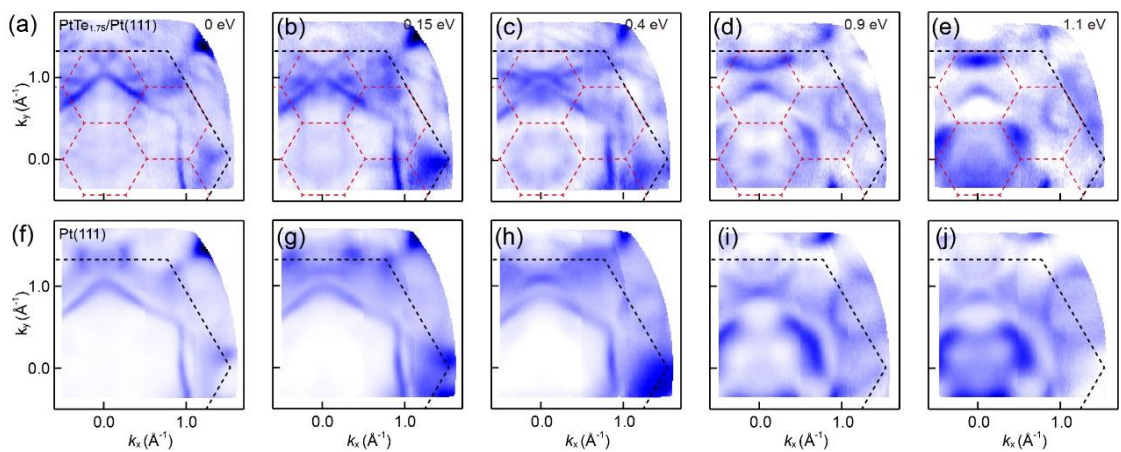

**FIG. 3: ARPES measurement of the CECs.** (a-e) ARPES intensity maps of PtTe$_{1.75}$/Pt(111) at different binding energies: 0, 0.15, 0.4, 0.9, and 1.1 eV, respectively. Black and red dashed lines indicate the BZs of Pt(111) and monolayer PtTe$_{1.75}$, respectively. (f-j) The same as (a-e) but for pristine Pt(111).

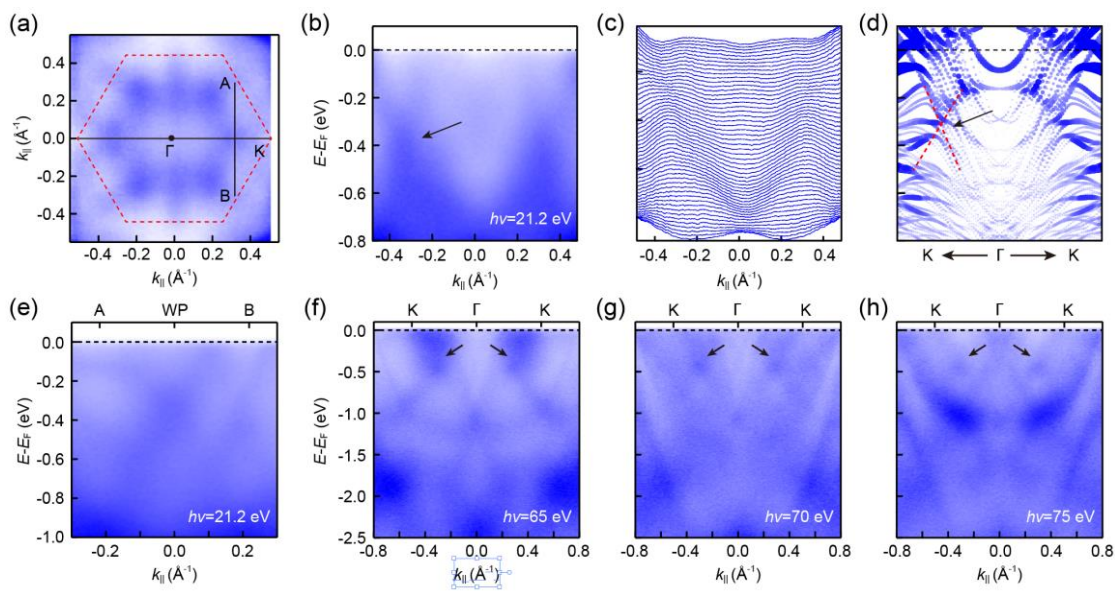

**FIG. 4: ARPES measurement of the band dispersion.** (a) Constant energy contour of monolayer PtTe$_{1.75}$ at 0.4 eV. The red dashed line indicates the first BZ of PtTe$_{1.75}$. (b) ARPES intensity plots along the K-Γ-K direction of PtTe$_{1.75}$/Pt(111). (c) Momentum distribution curves extracted from (b).

(d) DFT calculated band structure of PtTe$_{1.75}$/Pt(111) along the K-Γ-K direction. (e) ARPES intensity plots along Cut AB, as indicated in (a). (f-h) Photon energy dependent band dispersion of PtTe$_{1.75}$/Pt(111). Black arrows indicate the Weyl cones, which shows negligible dispersion with photon energy.

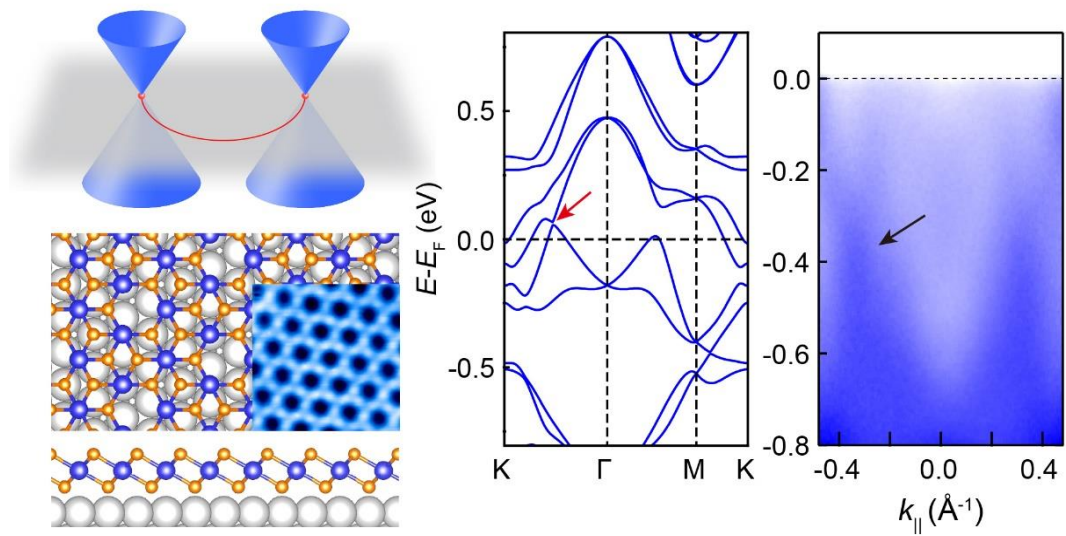

**TOC Graphic**